# Protected Long-Distance Guiding of Hypersound Underneath a Nano-Corrugated Surface

Dmytro D. Yaremkevich[1], Alexey V. Scherbakov[1,2], Serhii M. Kukhtaruk[1,3], Tetiana L. Linnik[3], Nikolay E. Khokhlov[2], Felix Godejohann[1], Olga A. Dyatlova[1], Achim Nadzeyka[4], Debi P. Pattnaik[5], Mu Wang[5], Syamashree Roy[5], Richard P. Campion[5], Andrew W. Rushforth[5], Vitalyi E. Gusev[6], Andrey V. Akimov[5], and Manfred Bayer[1,2]

[1]*Experimentelle Physik 2, Technische Universität Dortmund, Otto-Hahn-Str. 4a, 44227 Dortmund, Germany.*

[2]*Ioffe Institute, Politekhnycheskaya 26, 194021 St. Petersburg, Russia.*

[3]*Department of Theoretical Physics, V. E. Lashkaryov Institute of Semiconductor Physics, Pr. Nauky 41, 03028 Kyiv, Ukraine.*

[4]*Raith GmbH, Konrad-Adenauer-Allee 8, 44263 Dortmund, Germany*

[5]*School of Physics and Astronomy, University of Nottingham, Nottingham NG7 2RD, United Kingdom.*

[6]*LAUM, CNRS UMR 6613, Le Mans Université, 72085 Le Mans, France.*

## Abstract

**Within a new paradigm for communications on the nanoscale, high-frequency surface acoustic waves are becoming effective data carrier and encoder. On-chip communications require acoustic wave propagation along nano-corrugated surfaces which strongly scatter traditional Rayleigh waves. Here we propose the delivery of information using subsurface acoustic waves with hypersound frequencies ~20 GHz, which is a nanoscale analogue of subsurface sound waves in the ocean. A bunch of subsurface hypersound modes is generated by pulsed optical excitation in a multilayer semiconductor structure with a metallic nanograting on top. The guided hypersound modes propagate coherently beneath the nanograting, retaining the surface imprinted information, on a distance of more than 50 μm which essentially exceeds the propagation length of Rayleigh waves. The concept is suitable for interfacing single photon emitters, such as buried quantum dots, carrying coherent spin excitations in magnonic devices, and encoding the signals for optical communications at the nanoscale.**

Acoustic waves with terahertz (THz) and gigahertz (GHz) frequencies, which are often referred to as hypersound, are attractive for applications in quantum computing, sensing, and communications.[1-4]. A major advantage is the nanometer wavelength of hypersound which, in contrast to photons, is comparable with the size of single-electron and single-photon devices. Surface acoustic waves (SAWs) of hypersound frequencies, being the high-frequency counterpart of traditional MHz SAWs, have been implemented successfully in quantum technologies as an instrument to control single electron,[5,6] photonic,[7,8] and spintronic[9-11] devices. It has been proposed to use hypersound to connect different nanoobjects (*e.g.* qubits) on-chip in analogy with wires connecting the elements of traditional electronics.[12-15] For this, the propagation distance of the acoustic signal should be at least comparable with the distance between the source and receiver. SAWs with frequencies up to 50 GHz can propagate on flat surfaces over millimeter distances,[16] but nanodevice architectures on single chips inevitably require nanometer processing and surface corrugation upon which the hypersound diffracts, reducing the SAW's mean free path from millimeters to micrometers.[17]

In this article, we demonstrate experimentally how subsurface hypersound propagates from the source to the receiver on distances more than 50 μm underneath a nanopatterned metallic layer. The principle is schematically demonstrated in Fig. 1a where surface and subsurface waves are propagating along a nanostructured surface. Both waves are generated at point A and serve to deliver the information, *e.g.*, a spatially or temporally modulated signal, to point C. Traditional SAWs, *i.e.* Rayleigh-like waves, do not reach the destination due to diffraction and scattering at the nanoobject B located on the SAW's path.[18] Alternatively, subsurface waves continue to propagate under object B and do not feel any surface corrugation, similar to how a submarine does not feel the storm at the sea surface. This idea is a nanometer analogue for the delivery of acoustic signals using non-attenuating guided underwater sound waves.[19] To realize this concept in practice, the subsurface hypersound should propagate parallel to the surface with preserved coherence and without strong attenuation and leakage to the bulk. Reaching this goal requires the design of structures that support long propagating hypersound modes, and the development of methods to generate and detect such modes. In the present work, we achieve both goals. We demonstrate experimentally the propagation of coherent hypersound wavepackets, with central frequency ~20 GHz, on macroscopic distances which essentially exceed the mean free path of high-frequency Rayleigh waves at the corrugated surface.

**Results and Discussion**

The samples consist of lateral one-dimensional nanogratings (NGs) fabricated from metallic ($Fe_{0.81}Ga_{0.19}$) layers deposited on GaAs substrates. Two samples were studied in our experiment. In the first sample (see Fig. 1b) the metallic layer with a thickness of $l$=105 nm was deposited onto a GaAs/AlAs superlattice (SL) grown on a (001)-GaAs substrate. As we shall show, the elastic parameters of the SL allow the waveguiding effect for hypersound. In the second sample, the metallic layer with $l$=100 nm was deposited directly onto a GaAs substrate without a SL. The NGs with a lateral size of 25×100 μm² consist of grooves of 25 nm depth milled parallel to the [010] crystallographic direction of the GaAs substrate. The grooves and stripes have the same width $a$, and the corresponding period $d$ in the two gratings has values $d$=200 and 150 nm, respectively. The scanning electron microscope image of the NG with $d$=200 nm is shown in Fig. 1c.

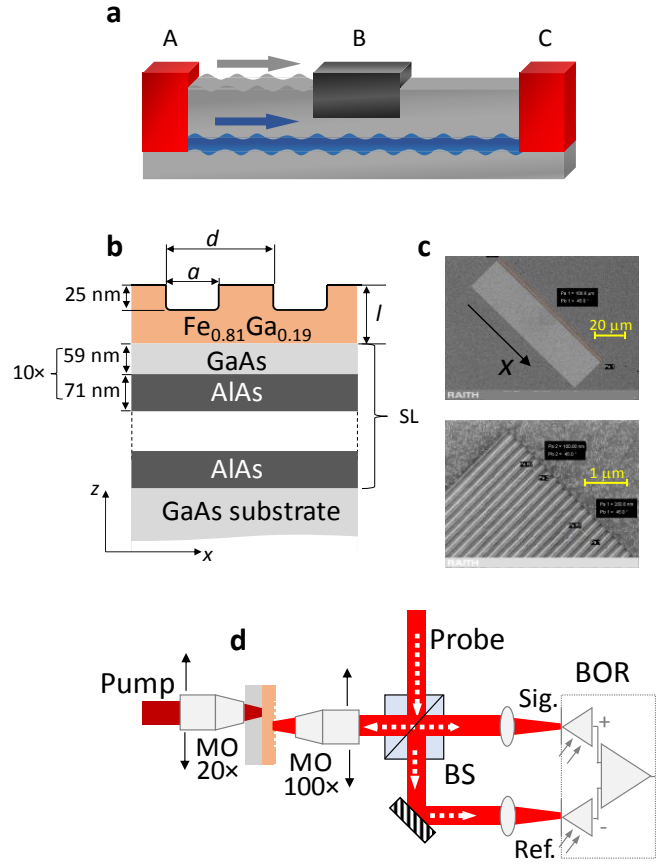

**Figure 1. Experimental setup. a** Basic concept: surface waves travelling from A to C are scattered by object B, while subsurface waves freely propagate beneath the corrugated surface. **b** Scheme of the sample. **c** SEM images of the nanograting. **d** Schematic of the experimental setup for measuring phonon propagation with the pump and probe spots separated in space.

Hypersound waves are excited and detected optically using a picosecond ultrasonic pump-probe technique.[20,21] The schematic of the experiment is shown in Fig. 1d. The pump beam from the femtosecond laser is sent in from the side of the GaAs substrate, which is transparent for the laser wavelength 1050 nm, and focused onto the metallic film on a Gaussian spot with radius $\sigma$=1 µm. The probe pulses from another laser with wavelength 780 nm are focused on a spot with a radius of 0.5 µm on the metallic grating. The femtosecond pump pulse induces instantaneously thermal stress in the metallic layer, leading to the generation of the acoustic wavepacket with hypersound frequencies up to 100 GHz.[20] In the presence of the NG, the generated hypersound propagates along the *x*-direction parallel to the surface and has wavevector components:

$$q_x = q_P + n\, 2\pi/d \qquad n=0;\ 1;\ 2\ldots \qquad (1)$$

where *x* is the [100] direction perpendicular to the grooves (see Fig. 1b), and $q_P \ll 2\pi/d$ is the wavevector limited by the size of the pump laser spot (see Methods). The detected reflectivity probe signal $\Delta R(t)$ is sensitive to waves with the same $q_x$ given by Eq. (1). In the present experiments, we concentrate on measurements of the probe signals on distances up to 80 µm from the pump spot along the *x*-axis (*x*=0 corresponds to the center of the pump spot, *t*=0 corresponds to the time moment when the pump pulse hits the studied structure).

The measured signal $\Delta R(t)$ at *x*=0, *i.e.* at the point of spatial overlap of the pump and probe spots, in the grating with *d*=200 nm and the GaAs/AlAs SL subsurface layer is shown in the upper panel of Fig. 2a. The signal shows high-frequency oscillations sitting on a slowly varying background. The origin of the oscillations is elastic vibrations, *i.e.* coherent hypersound waves, while the slow background, shown by the red solid line, is known to be due to the modulation of the refractive index by hot electrons, thermal phonons and low-frequency acoustic motion of the surface excited by the optical pump pulse.[20,21] The inset in Fig. 2a shows the fast Fourier transform (FFT) of the measured signal within three spectral bands: B1 (10 – 15 GHz); B2 (15 – 20 GHz); and B3 (25 –30 GHz). The low-frequency band B1 includes an intense narrow spectral line at $f_{R1}$=12.1 GHz which corresponds to the first-order Rayleigh-like mode observed earlier in experiments with NGs.[22-26] The other spectral bands include several surface modes, quasi-surface (skimming) modes, and bulk modes, which are commonly seen in experiments with gratings when the pump and probe pulses overlap.[17,25,26] The lower panel shows the temporal signals after filtering in the different spectral bands. The signal in B1 has the longest lifetime while the oscillations in the bands B2 and B3 decay much more quickly.

Figure 2b shows the results for $\Delta R(t)$ measured when the pump and probe spots are separated in space by 10 µm along the *x*-direction. In this case, only hypersound waves propagating along the surface perpendicular to the grooves are detected. From the raw signal (Fig. 2a) it is seen that the slowly varying background is absent which is explained by localization of thermal phonons and hot electrons inside the pump spot. The measured signals $\Delta R(t)$ show oscillations that sit on a background from the pulse with a duration of ~1 ns (shown by the solid red line). This nanosecond pulse corresponds to the generated acoustic signal with $q_x \leq q_P$ ($n$=0),[27] which is excited also in a plain film without NG, and will not be discussed further. The filtered signals (lower panel) and their FFTs (inset) differ significantly from the case $x$=0: there are several hypersound modes in B1 and the spectral line in B2 centered at $f_W$~17 GHz has an intensity comparable with the modes in B1. There is a weak contribution from high-frequency modes in B3. The signals in the different spectral bands appear at different delay times $t_i = x/s_i$ ($s_i$ is the group velocity of the wavepacket in the corresponding frequency band) relative to the pump pulse which suggests different velocities $s_i$ of the corresponding hypersound modes. All reflectivity traces show beatings in the filtered signals which points towards multiple mode composition of the hypersound waves in each spectral band.

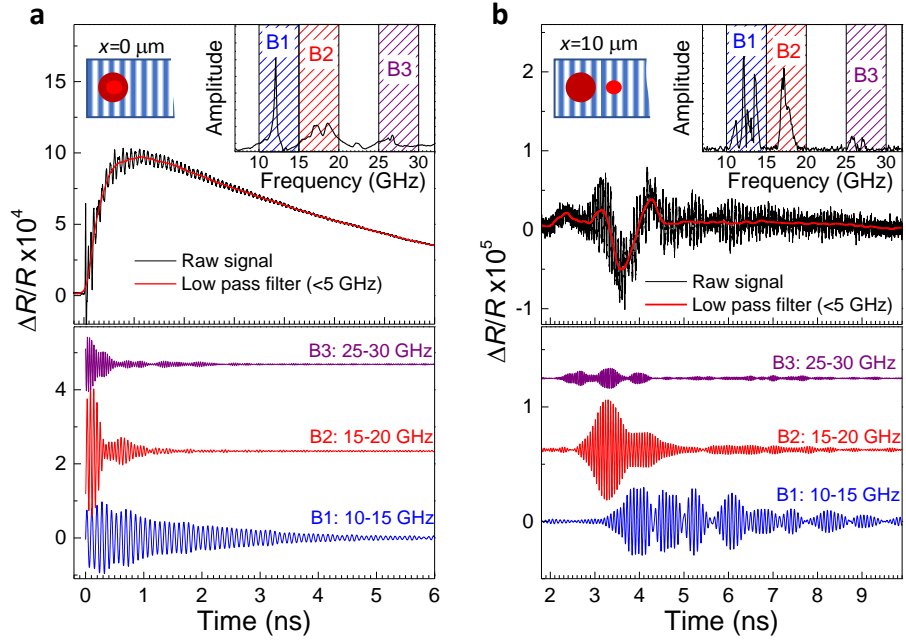

**Figure 2. Experimental signals. a**. Upper panel: signal detected when the pump and probe spots overlap in space; the inset is the fast Fourier transform (FFT) of the temporal signal. Lower panel: temporal traces after filtering of the measured signal from the upper panel in three frequency bands marked in the inset. **b.** Same as **a** but measured when the pump and probe spots are separated on the surface by 10 μm.

Figure 3a shows the evolution of the FFTs with the increase of the distance to $x$=50 μm. The normalized spectra show that only the hypersound modes in B2 with frequencies $f_W$~17 GHz remain in the wavepacket at large distances. We shall call these long propagating modes *waveguide modes* (W-modes). The reason for this name will become clear after identifying their polarization and spatial distribution. All other modes, including the traditional Rayleigh-like modes in the B1 spectral band (further referred to as R-modes) decay much faster. Figure 3b compares the decays of the R- and W- modes with distance. Blue circles show the dependence of the spectral amplitude of the first-order Rayleigh mode (R1) at $f_{R1}$=12.1 GHz. This mode is dominating in the signal measured at $x$=0 and demonstrates the longest propagation among the R-modes. The mean free path for the R1-mode is $\bar{l}_{R1} = 7.2$ μm, so that it is hardly observable at $x$=30 μm. In contrast, the W-modes have a much larger mean free path $\bar{l}_W$. The distance dependence of their mean spectral amplitude averaged over the B2 frequency band, shown by red squares, demonstrates a non-monotonic decay and is almost constant at $x$>30 μm. The W-modes are uniquely detected in the time domain at $x$=50 μm and larger (our measurements are limited by the NG length). Figure 3c shows the filtered signals for the W-modes (B2 spectral band) at several distances from the pump spot. The signals possess beatings and their exact shape depends on $x$. This indicates that the signal in this spectral band consists of several hypersound modes propagating at different velocities. The inset in Fig. 3c shows the linear dependence of the arrival time of the W-modes (determined from the center of the respective wavepackets) on the coordinate *x*. From this dependence, we obtain the mean sound velocity $\bar{s}_W = 3170$ m s$^{-1}$ which is slower than the sound velocities for transverse (TA) sound in the [100] direction in metallic (Fe,Ga) ($s_{FeGa}$=3955 ms$^{-1}$) and GaAs ($s_{GaAs}$=3346 ms$^{-1}$), but is essentially faster than the measured speed of sound for the R1-mode in the studied sample ($s_{R1}$=1950 ms$^{-1}$).

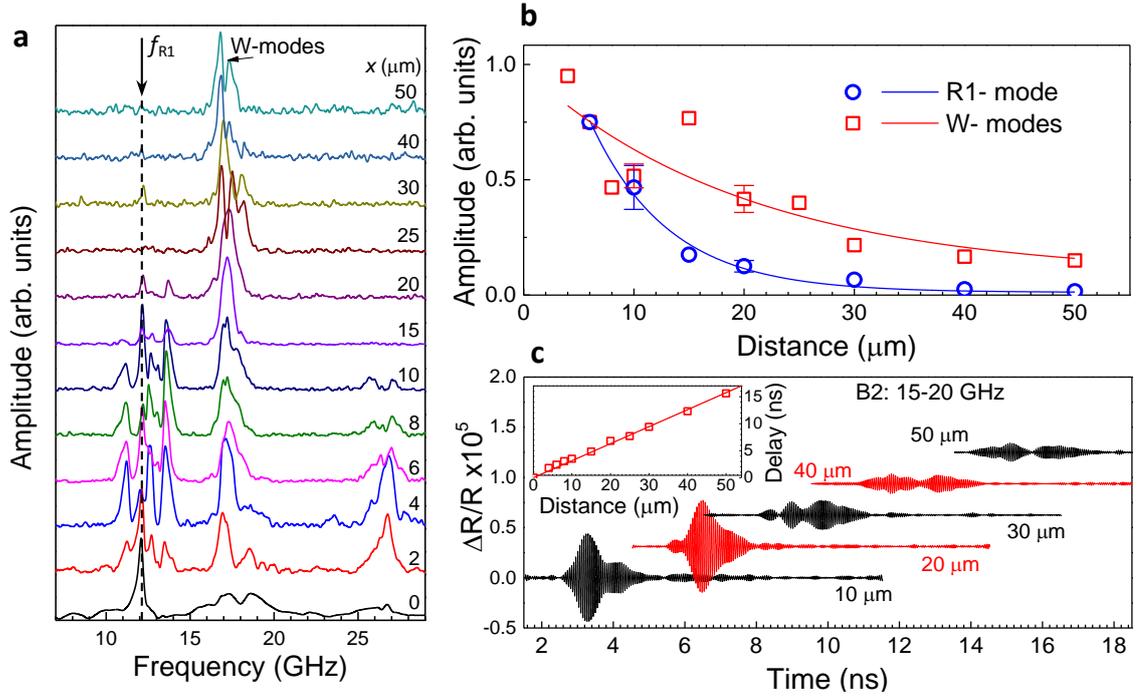

**Figure 3. Dependence on the distance between source and detector. a** FFTs obtained from the signals measured at various distances $x$ between the pump and probe spots. Each spectrum is normalized to its maximum spectral amplitude. **b** Dependences of the spectral amplitude of the filtered signals for the Rayleigh wave (R1 mode) and the mean spectral density for the waveguide modes (W-modes) on the distance between the pump and probe spots. Solid lines show the fits of the experimental data by an exponential decay, which serves to estimate the averaged mean free path for the propagating modes. The data and fit curves are normalized for clear presentation. **c** Temporal signals after filtering in the frequency band B2 for various distances between the pump and probe spots; the inset shows the dependence of the arrival time of the bunch around 17 GHz on the distance.

Another nanograting with *d*=150 nm deposited on the GaAs/AlAs SL sublayer shows similar results, but metallic layers deposited onto a GaAs substrate without SL do not show any modes, which propagate on distances larger than 10 μm in any spectral band. The experimental results and the calculated values of sound group velocities $s_i$, estimated mean free paths $\bar{l}_i$, and maximum distances $l_i^{max}$, at which hypersound is clearly seen in the FFTs above the noise level, are presented in the Supporting Information for all measured samples. Thus, we conclude that the SL subsurface layer is essential to observe hypersound modes propagating on long distances.

To understand the difference in the propagation length for Rayleigh-like and W-hypersound modes, we identify the origins of the observed spectral lines in the FFTs in the phonon dispersion curves. Figure 4a shows these dispersion curves, $f(k_x)$, for the sample with the *d*= 200 nm nanograting in the metallic Fe$_{0.81}$Ga$_{0.19}$ layer, deposited on the GaAs/AlAs SL. Here $k_x$ is the Bloch wavevector and $q_x = k_x + n\,2\pi/d$. Only modes with atom displacements **u** (*x, z*) in the sagittal plane (*xz*-plane in Fig. 1b) are considered. The dispersion curves are shown in the range near the center of the folded Brillouin zone for the frequency range around *n*=1. The calculated dispersions for the plain structure without NG (see S2 in Supporting Information) help us to identify the origins of the modes. We attribute the dispersion branches R1 and R1* to the lowest first-order Rayleigh-like mode. In the periodic structure with NG, this mode is split into the symmetric (R1) and antisymmetric (R1*) modes with even and odd distributions of the *z*-component of the displacement relative to the center of the NG grove, with a spectral gap at $k_x$=0. The spatial distributions of these modes are concentrated in the metallic layer as demonstrated in Fig. 4b. Both R1 and R1* modes are observed in the experiments for relatively small distances 2<*x*<15 μm at $f_{R1}$=12.1 GHz and $f_{R1*}$=11.2 GHz, respectively (for example, see the inset in Fig. 2b). The modes R2 and R2* are the second-order Rayleigh-like modes, often called Sezawa-like modes.[28] Similar to the R1-modes, these modes are concentrated in the (Fe,Ga) layer. The excellent agreement with the calculated frequencies $f_i$ and sound velocities $s_i$ allows us to identify unambiguously the symmetric (R1 and R2) and antisymmetric (R1* and R2*) propagating Rayleigh modes in the experimental spectra.

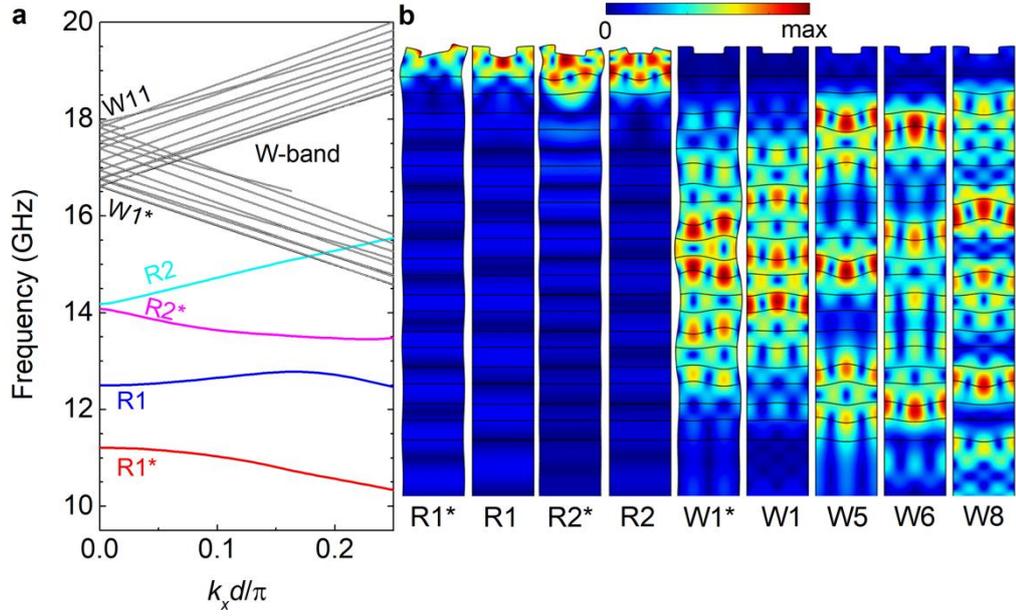

**Figure 4. Dispersions and spatial distributions of the hypersound modes. a.** Calculated dispersion curves for hypersound modes propagating along the surface in the structure with the $Fe_{0.81}Ga_{0.19}$ nanograting with $d$=200 nm deposited on the GaAs/AlAs superlattice. **b.** Spatial distribution of the absolute value of displacement vector (exaggerated for clarity) for the Rayleigh waves (R1 and R1*) localized in the metallic layer with the nanograting, and for 5 waveguiding (W) phonon modes localized in the GaAs/AlAs superlattice.

The most important result revealed by the dispersion curves is the existence of the W-modes with linear dispersion branches that are degenerate at $k_x$=0 and $f_W$~17 GHz, *i.e.*, with Dirac-cone type dispersions. The striking difference between the W- and R-modes is their spatial distribution as shown in Fig. 4b. While the R-modes are concentrated in the metallic layer, the W-modes are localized in the sublayer with the GaAs/AlAs SL, well below the NG. The W-modes have an analogy with the Lamb modes in plates and behave as waveguide modes. In total, there are 11 pairs of W-modes localized in the SL (see section S3 in Supporting Information for details). Their frequencies and group velocities cover the ranges $f_W$=16.5-18 GHz and $s_W$=2800-3600 ms$^{-1}$, respectively. The excellent agreement between calculated and measured frequencies and their velocities leads us to the conclusion that the waveguide W-modes are those with the long propagation lengths in our samples.

The dispersion curves and spatial distributions for the sample without a SL subsurface layer show that R-modes exist there with dispersion relations very similar to the considered case with a SL, but W-modes are not present. This theoretical result confirms that the observation of W-modes is governed exclusively by the presence of the GaAs/AlAs SL which acts as a multichannel waveguide. The same conclusion can be drawn by comparing the signals measured on the samples with and without a SL sublayer.

The exciting property of the W-modes is their ability to propagate parallel to the surface for long distances. This is explained by the localization of the W-modes in the SL layer as shown in Fig. 4b. The (Fe,Ga)/SL and SL/GaAs interfaces play the role of confining borders for Lamb-like modes [29] in the SL sublayer. The structure could be viewed as a waveguiding plate loaded on top by a layer of finite thickness and at the bottom by the infinite substrate. In comparison with a Lamb plate in air or water, the structure is asymmetric both geometrically and in terms of the loading materials. In contrast to Rayleigh waves at a plain surface, the corrugation results in the leakage of both R- and W-modes to the bulk.[18] However, the calculations show that the losses into the bulk modes are much smaller for the guided W-modes than for the R-modes. The calculated quality factor, $Q$, reaches ~$10^5$ for the W-modes, but does not exceed ~$10^3$ for the R-modes (see Table S2 in the Supporting Information). This difference is explicitly demonstrated in the Supporting Video 1, which animates the calculated propagation of the excited hypersound in the studied sample and visualizes the emissions from R-modes and W-modes into bulk waves. When the R-modes and W-modes are sufficiently spatially separated to distinguish these two emissions in the animation (after 4.5-ns delay from the optical excitation), the strong leakage of R-modes to the bulk is clearly observed, while the weak leakage of W-modes is hardly visible. The linear dispersion of W-modes (see Fig. 4a), in addition to their high $Q$ factors, also indicates that the grating, which causes the excitation of the hypersound waveguide modes, has a small influence on their propagation. Thus, the much larger broadening of the R-modes' wavepacket in comparison with the wavepacket of the W-modes is clearly visible already at small $x$ (compare the filtered signals for the B1 and B2 bands in Fig. 2b).

The waveguiding observed experimentally is a multimodal effect. The bunch of propagating W-modes covers the frequency range from 16 to 18 GHz (see the FFTs in Fig. 3a). The temporal signals and their FFTs allow us to distinguish several modes propagating with different velocities. As a result, we observe temporal beatings in the signals presented in Fig. 3c and a non-monotonic dependence of the mean spectral density on the coordinate $x$ in Fig. 3b. To illustrate the multimodal character of the waveguiding effect, we present a simulation of the transient reflectivity signal induced by a propagating multimode wavepacket in the Supporting Video 2. Two signals from this simulation are presented together with the experimental signals for $x$=20 and 40 μm in Fig. 5. In the simulations, we did not aim for a perfect agreement with the experiment and limited the signal in the B2 band to the sum of 4 W-modes that are simultaneously excited at the excitation spot with the same amplitudes, spatial distributions, and initial phases. Their frequencies $f_w$= 16.532, 17.026, 17.223, and 17.455 GHz and corresponding velocities $s_w$= 3266, 3310, 3591, and 2881 ms$^{-1}$ are taken from the calculated dispersion curves in Fig. 4a and correspond to $k_x = 0.01\,\pi/d$. It is seen that the arrival time of the hypersound bunch and its temporal width are in good agreement with the experiment. The simulated spatial-temporal evolution of the signal shows also pronounced beatings, which agrees with the experimental observations and supports the conclusion about the multimodal character of the waveguiding effect.

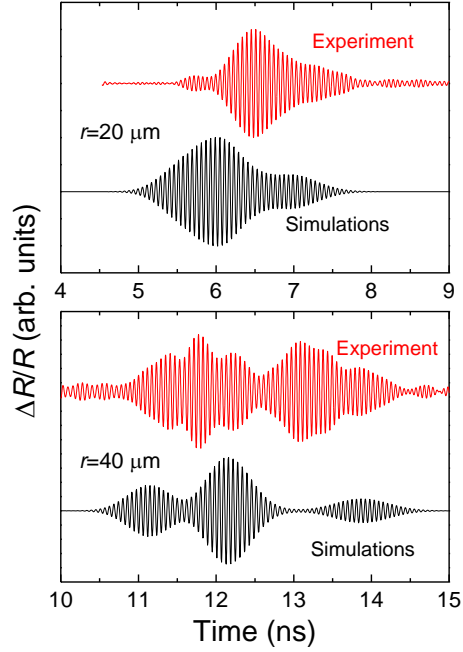

**Figure 5. W-mode wavepackets.** Measured (red) and simulated (black) phonon wavepackets in the B2 spectral band for two propagation distances: $x$=20 μm (upper panel) and 40 μm (lower panel). In the simulations 4 W-modes with frequencies $f_w$= 16.532, 17.026, 17.223 , and 17.455 GHz and velocities $s_w$= 3266, 3310, 3591, and 2881 ms$^{-1}$ are summed.

**Conclusions**

In summary, we have realized experimentally generation and detection of acoustic waves with frequency up to 20 GHz which propagate underneath a corrugated surface. These hypersonic waves propagate on a distance of more than 50 μm and lose their energy to bulk phonons much less efficiently than Rayleigh waves which propagate only on several micrometers due to scattering on the nanostructured surfaces.

For prospective applications in communications, it is important that the subsurface W-modes faithfully carry the surface morphology imprinted at the point of excitation, which must then be read-out at the point of detection. In the particular case described in the present experiments, the morphology is carried by the generated and detected hypersound wavevector $q_x \approx 2\pi/d$, defined by the NG period. In the general case, the excitation and detection points could include nano-phononic cavities [30] coupled with the subsurface waveguide. A further advantage over traditional Rayleigh-like modes may be taken from the bunched character of the subsurface W-modes, the number of which is controlled by the width of the waveguiding layer. This makes possible using the same subsurface waveguide for encoding information at different acoustic frequencies with the same $q_x$ by resonant driving.[31]

The possibility to deliver information by hypersound well beneath corrugated surfaces on macroscopic distances has a direct perspective to control a single photon from buried quantum nanoobjects [8,32,33] which quantum efficiency decreases strongly when located near the surface. Subsurface hypersound may be used to control the emission from a Bose condensate in polariton microcavities located at micrometer distance from the surface.[34] The technique may be applied to control exciton resonances in two dimensional van der Waals nanolayers, which require passivation for high quantum efficiency.[35,36] The (Fe,Ga) metallic layer used in our experiment is a ferromagnet, known as Galfenol.[37] It possesses a strong magnon-phonon coupling which allows the conversion of magnon excitations to hypersound and vice versa.[38] In magnonic devices, the described concept of a nanostructured ferromagnetic layer deposited on a hypersound waveguide can be used for transmitting coherent spin excitations.[39,40] The preliminary measurements show that despite of dominating localization of the W-modes in the SL, their tail located in the Galfenol layer has sufficiently large amplitude for carrying coherent magnons of the corresponding frequencies for distances exceeding 50 μm. This effect, however, is out of the scope of the present paper and will be addressed elsewhere.

It is appealing to excite the guided sub-surface modes electrically without using bulky ultrafast lasers. This method, however, should be different from standard techniques for generating SAWs with interdigital transducers. It is necessary to excite a bulk hypersonic wave, which propagates perpendicular to the surface and carries the information encoded at the surface (*e.g.* by a grating) in terms of its lateral amplitude and phase modulation. When incident on the undersurface waveguide, this laterally modulated wave transfers the encoded information to the W-modes. The electrical generation of bulk hypersonic waves at frequencies higher than 10 GHz is challenging. Recently such generation of hypersound with frequency up to 20 GHz has been demonstrated in Ref. [41] with piezoelectric ZnO transducers; application of such electrical technique to the corrugated surface could lead for wider applications of subsurface hypersound.

**Methods**

**Samples Production.**

The studied samples were epitaxially grown on commercial GaAs substrates [(100)-semi-insulating GaAs]. The superlattice was produced by molecular beam epitaxy and the $Fe_{0.81}Ga_{0.19}$ layers were deposited by magnetron sputtering. The nanogratings were fabricated through focused ion beam milling (Raith VELION) with both 100-nm and 75-nm lines and spaces over an area of 25 µm x 100 µm. To ensure high beam resolution during the patterning of the gratings a low Ga+ beam current of 22 pA was applied at 35 keV beam energy. The milling dose was set to 0.3 nC/µm$^2$ resulting in the groove depth of 25 nm.

**Time-Resolved Pump-Probe Measurements**.

The pump-probe scheme was realized using two mode-locked Erbium-doped ring fiber lasers (TOPTICA FemtoFiber Ultra 1050 and FemtoFiber Ultra 780). The lasers generate pulses of 150-fs duration with a repetition rate of 80 MHz at wavelengths of 1050 nm (pump pulses) and 780 nm (probe pulses). The pump beam was focused by a microscope objective (20×; N.A.=0.4) to the backside of the metal layer through the GaAs substrate and GaAs/AlAs superlattice, which are transparent for the pump wavelength. The focused pump spot had a Gaussian intensity distribution $\sim \exp\left(-\frac{r^2}{2\sigma^2}\right)$, where $r$ is the distance from the center of the spot and $\sigma$=1 µm is the spot radius at $1/\sqrt{e}$ level. The maximal used pulsed excitation density was 8 mJ/cm$^2$. The probe beam was focused by another microscope objective (100×; N.A.=0.8) to the front side of the sample. The focused probe spot had a Gaussian distribution also with $\sigma$=0.5 µm and an energy density of 2 mJ/cm$^2$. The temporal resolution was achieved employing an asynchronous optical sampling (ASOPS) technique [42]. The pump and probe oscillators were locked with a frequency offset of 1600 Hz. In combination with the 80-MHz repetition rate, it allowed measurement of the time-resolved signals in a time window of 12.5 ns with a time resolution of <1 ps. The distance between the centers of the pump and probe spots was controlled with a precision of 0.05 µm. Two-dimensional spatial scanning across the nanograting surface confirmed the directional propagation of the hypersound wavepackets along the reciprocal lattice vector of the nanograting. At the considered propagation distances, the influence of diffraction on these acoustic beams is negligible and cannot be observed. The measured size of the hypersound beams and the *y*-distribution of the modes' amplitudes correspond to the size of the pump spot at the point of excitation. Only the low-frequency background Rayleigh wave [see the filtered (<5 GHz) signal in Fig. 2 (b)] is emitted in all surface directions and exhibits cylindrical symmetry.

**Transformation of Optically Generated Stress to Hypersound.**

The guided modes of interest belong to the system of acoustic eigenmodes of the sample, which are all spatially periodic because of the periodicity imposed by the nanograting (Floquet-Bloch theorem). Each eigenmode contains an infinite number of wavevectors: $q_x = k_x + n\,{2\pi}/{d}$, where $k_x$ is the Bloch vector, $d$ is the NG period and $n$ is an integer, but one spatial harmonic with a certain $n$ dominates. A particular acoustic eigenmode is excited with non-zero amplitude when the frequency spectrum of the photo-induced stress contains the frequency of the eigenmode and the lateral distribution of the stress contains the wavevector of this mode.[22] The intensity envelope of the fs- laser pulses applied in our experiments contains frequencies up to ~0.1 THz and, thus, covers the frequencies of the considered guided modes. In contrast, the spatial spectrum of the stress that is induced by the pump light focused on the interface of the SL and metallic layer (opposite to the open surface with the NG) is limited by the spatial pump intensity distribution. For a Gaussian distribution of the laser intensity along the $x$-axis $\sim \exp\left(-\frac{x^2}{2\sigma^2}\right)$, the spectrum of the wavevectors, $q_P$, generated by the photo-excited stress is proportional to $\exp\left(-\frac{q_P^2 \sigma^2}{2}\right)$. Due to the large size of the laser spot in comparison with the NG period, this spectrum does not contain components of $q_x$ with non-zero $n$. However, the generation of all acoustic eigenmodes is possible due to the matching of the photo-induced stress and the $n=0$ components of these eigenmodes.

**Calculation of Hypersound Dispersion and Spatial Distribution of Phonon Modes.**

We use COMSOL Multiphysics® for calculation of the phonon dispersion in our system. For this, we consider a NG unit cell and apply Floquet-Bloch periodic boundary conditions, *i.e.* $u_d = u_s \exp(-ik_x d)$, where $u_d$ and $u_s$ are the atom displacements at the destination and source boundaries, respectively, and $k_x$ is a Bloch wavenumber. At the open NG surface, we use free boundary conditions. For the simulation of the semi-infinite substrate, we use a perfectly matched layer (PML) at the backside of the substrate. The stiffness tensor components, $C_{kl}$, and the mass density, $\rho$, used for the calculations are: for Fe$_{0.81}$Ga$_{0.19}$ [32] $C_{11} = 209$ GPa, $C_{12}=156$ GPa, $C_{44} = 122$ GPa, $\rho = 7800$ kg m$^{-3}$; for GaAs $C_{11} = 119$ GPa, $C_{12}=53.8$ GPa, $C_{44} = 59.5$ GPa, $\rho = 5316$ kg m$^{-3}$; for AlAs $C_{11} = 119.9$ GPa, $C_{12}=57.5$ GPa, $C_{44} = 56.6$ GPa, $\rho = 3760$ kg m$^{-3}$. The dispersion and modes' spatial distributions are obtained by an eigenfrequency analysis. Most of the solutions are shown in Fig. 4a. There are also solutions arising due to the finite size of the simulated substrate, but they are not related to the experimentally observed effects and are, thus, ignored. To select the modes localized in the NG and SL, we used the approach described elsewhere.[43] We characterize the localization degree with the quality factor $Q$ and parameter $p$ defined by:

$$p = \frac{<F_e>_L}{<F_e>_S}, \qquad (2)$$

where $F_e$ is the elastic free energy density and the averaging is performed over two areas: the L-area includes the Fe$_{0.81}$Ga$_{0.19}$ film and the SL; the S-area corresponds to the PML. Only W-modes with $p \gg 1$ propagate for a long distance. The value of $p$ depends on $q_x$. The calculations show that for the $Fe_{0.81}Ga_{0.19}$ film $p>100$ and $Q > 5 \cdot 10^3$ for 4 W-modes at $q_x \approx 2\pi/d$ (see S3 in Supporting information). These 4 W-modes from the total of 11 pairs are considered as the best candidates for long propagating subsurface modes in our experiments. The existence of the W-modes with long propagation length is confirmed by a semi-analytical consideration which is described in Supporting Information (section S2).

**Acknowledgements**

The work was supported by the Bundesministerium fur Bildung und Forschung through the project VIP+ "Nanomagnetron" and by the Deutsche Forschungsgemeinschaft in the frame of the Collaborative Research Center TRR 142 (project A06 "Tailored ultrafast acoustics for light emission modulation"). The cooperation between TU Dortmund, the Lashkaryov Institute, and the Ioffe Institute was supported by the Volkswagen Foundation (Grant No. 97758).

**Supporting Information.**

Experimental and theoretical data for all the studied samples; elastic equations and their semianalytical solutions for an unpatterned metallic film on a GaAs substrate and on a superlattice; spatial distributions and main parameters of the W-modes for the grating with 200-nm period.

https://drive.google.com/file/d/1YxZUZ2Ty5eWcbblA0X-7wLSrR0I31Tqj/view?usp=sharing

**Supporting Videos.**

Supporting Video 1 demonstrates the spatial-temporal modeling (COMSOL Multiphysics®) of the hypersound propagation in the structure with a GaAs/AlAs superlattice and a $Fe_{0.81}Ga_{0.19}$ nanograting with period $d$=200 nm after excitation by the pump pulse. The color map illustrates the time-derivative of the $z$-component of the displacement vector. The black horizontal lines show the boundaries between the different materials of the structure.

https://drive.google.com/file/d/1eNNdF3EfA7C9agjeaK2O_YNCegqowMHE/view?usp=sharing

Supporting Video 2 illustrates the temporal evolution of the guided wavepacket consisting of 4 W-modes with frequencies $f_w$= 16.532, 17.026, 17.223, and 17.455 GHz and corresponding velocities $s_w$= 3266, 3310, 3591, and 2881 ms$^{-1}$, taken from the calculated dispersions. The modes are simultaneously excited at the excitation spot with the same amplitudes, spatial distributions, and initial phases.

https://drive.google.com/file/d/1nusx51yGsP3mjwyvf9cM4L8fa2snZShS/view?usp=sharing